\documentstyle[new_cite]{mn}

\input{psfig}

\title{The unusual host galaxy of the BL Lac object PKS 1413+135}
\author[]
  {G.~Lamer,$^1$\thanks{E-mail: gl@astro.soton.ac.uk} 
  A.M.~Newsam,$^{1,2}$, and I.M.~M$^{\rm c}$Hardy$^1$ \\
  $^1$Department of Physics and Astronomy, The University,
      Southampton, SO17 1BJ\\
  $^2$Astrophysics Research Institute, Liverpool John Moores University, 
      L3 3AF}
\date{Accepted. Received}
\pagerange{\pageref{firstpage}--\pageref{lastpage}}
\pubyear{1998}

\pagerange{\pageref{firstpage}--\pageref{lastpage}}
\pubyear{1998}

\begin{document}

\maketitle

\label{firstpage}

\begin{abstract}
The BL Lacertae object PKS~1413+135 is associated with a disk dominated
galaxy which  heavily absorbs the BL Lac nucleus at optical and
X-ray wavelengths. It has been argued whether this galaxy is
actually the host galaxy of PKS~1413+135 or whether the BL Lac is
a background QSO, gravitationally lensed by the apparent host galaxy.
We have obtained deep high resolution H-band images of this 
unusual BL Lac object using the UKIRT IRCAM3. 
Our observations show that the BL Lac nucleus is centered  
within $\leq 0.05$ arcsec of the galaxy. Based on this result we assess
the probability for the lensing scenario and come to the conclusion
that the disk galaxy is indeed the host  
of PKS~1413+135. The galaxy shows  peanut-shaped isophotes,
suggesting the presence of a central bar which is a common feature of AGN.

\end{abstract}


\section{Introduction}
The flat spectrum radio source PKS 1413+135 was classified as a BL Lac
object on the basis of its lack of emission lines, rapid variability,
and strong and variable polarization \cite{bregman}.   
Deep optical imaging   of PKS~1413+135 
\cite{imh91}  showed its  host galaxy  to be disk
dominated. 
This finding is at variance with the unification of
BL Lac objects with Fanaroff-Riley type I (FR I) radio galaxies,
which have elliptical host galaxies. However, recently the discovery of 
a FR I galaxy with disk dominated host was reported by \scite{Ledlow}.
A comparison of this radio galaxy with PKS 1413+135 is given in 
section \ref{Disc}. 
 
The very red BL Lac  nucleus, which is not detectable  at optical
wavelengths, and photoelectric absorption in X-rays by a column of
$N_{\rm H}> 2\cdot 10^{22} {\rm cm}^{-2}$  \cite{Stocke92} led to the 
conclusion
that the BL Lac is heavily absorbed by interstellar matter of the host
galaxy which is viewed nearly edge-on. This is supported by 
the detection of a prominent dust lane in
V band images obtained with the {\em Hubble Space Telescope}
\cite{imh94}. 
An AGN within a gas rich galaxy would be expected to heat
any surrounding matter and  emit reprocessed radiation as a thermal
IR continuum or emission lines. Neither type of reprocessed
emission has been observed in PKS~1413+135.
Furthermore the classification as a BL Lac object implies that the 
AGN is oriented with the jet axis towards the line of sight. 
Thus the jet axis would be oriented perpendicular to the rotation axis
of the galactic disk.
To avoid these problems, \scite{Stocke92} have suggested that 
the BL Lac might be a background quasar
whose compact nuclear emission is amplified by gravitational
microlensing.

If the BL Lac actually lives in the apparent host galaxy, we would
expect it to lie exactly in the centre of the galaxy. 
However, if it is a background
microlensed quasar, 
it would be  unlikely to find the  point source very closely aligned with 
the foreground galaxy (e.g. \ncite{Merrifield}).   
We would also expect multiple images of the
background source due to gravitational macrolensing. 

In order to detect both the galaxy and the core, which is very red,
near infrared images are required. \scite{Stocke92} present the
best such images to date, in which they detect the host galaxy in 
H-band out to nearly 3 arcsec from the core. Although they do not
model the host galaxy and core to determine relative positions, or the
photometric parameters of the host galaxy, they state that the
position of the point source coincides with the centroids of the 
 elliptical isophotes of the galaxy within 0.1 arcsec .
However, \scite{Stocke92} still preferred the lensed 
interpretation, 
arguing that the lack of multiple lensed images could be due
to the absence of a significant bulge in this galaxy.

Here we present significantly deeper H-band images than that of Stocke
et al., in which the galaxy can easily be detected to at least 5 arcsec
radius from the core. We also clearly detect the object 6 arcsec east 
of the galaxy which is not detected in previous near 
infrared images.

We show that the core is centroided within the galaxy to at least
0.05 arcsec, and probably better, confirming that the core is very
unlikely to  be a 
background lensed QSO. We also show that the outer isophotes of the  
host galaxy have a boxy appearance indicative of a central bar, which
are commonly found in AGN.
Deviations from a simple disc form were only hinted at in our HST
V-band observations \cite{imh94}, but they are much better defined here.

\section{Observations and data analysis} 

High resolution H-band observations of PKS~1413+135 have been
performed with the {\em United Kingdom Infrared Telescope}
(UKIRT) equipped with  IRCAM3 
and a tip-tilt adaptive optics system  during two nights in January/February
1998. As the effective seeing during the observations 
was as good as 0.5-0.7 arcsec (FWHM), we used the 2$\times$  magnifier 
in order to achieve a sampling of 0.143 arcsec pixel$^{-1}$.
The resulting field of view is 37$\times$37 arcsec. 
Between exposures the source position was moved on the detector within
a 3$\times$3 grid with 4 arcsec spacings.
As no other bright point source is visible within the field of view of
PKS 1413+135, several images of a nearby star were taken in order to 
measure the point spread function of the telesope and adaptive optics
system. As the performance of the tip-tilt system likely depends on
the magnitude, distance, and position angle of the guide star
relative to the object, we have chosen the PSF reference star and 
guide star pair so that the parameters of this guide star were 
similar to that used for PKS 1413+135 itself.    

\begin{table}
\centering
\caption{\label{obs_log} H-band observations}
\begin{tabular}{ @{}lll@{} }
date & total exposure & eff. seeing \\[10pt]
31-Jan-1998 &  2560 s &   0.57 arcsec \\
01-Feb-1998 &  3180 s &   0.66 arcsec \\
\end{tabular}
\end{table}

The exposures of each night were recentered and coadded after dark
subtraction, bad pixel masking, and flat fielding.  
The flat fields were constructed by median combining the science 
images after masking out bright sources.

A two-dimensional fitting routine was used to derive the parameters 
of the core and host galaxy
components of PKS 1413+135.
Two models for the host galaxy were used for fitting:
An  exponential disk model with the radial intensity profile 

\[ I(r)=\exp(-1.6783 \cdot \frac{r}{R_{\rm s}}) \]

and a de Vaucouleurs model with the profile

\[  I(r)=\exp(-7.67 \cdot (\frac{r}{R_{\rm s}})^{1/4}-1 ) \] 

with $R_{\rm s}$ being the major axis scale length of the galaxy.
Galaxy and point source images  where simulated using the 
{\tt IRAF} task {\tt mkobjects} and then convolved with the normalized image
of the PSF star. The core and galaxy images where then added together 
and the resulting model image was compared with the H-band image of 
PKS 1413-135. The complete model is defined by the following set
of parameters:

\begin{tabular}{ll}
$m_{\rm gal}$:  & magnitude of the galaxy\\
$R_{\rm s}$:    & scale length the of galaxy\\
$e_{\rm gal}$:  & ellipticity\\ 
$\Phi_{\rm gal}$: & position angle\\
$x_{\rm gal}$:  & x coordinate of galaxy center\\ 
$y_{\rm gal}$:  & y coordinate of galaxy center\\
$m_{\rm nuc}$ : & magnitude of the nucleus\\
$x_{\rm nuc}$: & x coordinate of nucleus\\ 
$y_{\rm nuc}$: & y coordinate of nucleus\\
\end{tabular} 
 
The Marquard-Levenberg fitting algorithm was used to optimize these parameters 
simultaneously. The best fitting solution proved to be stable and
independent of the starting parameters.

We have adopted the  parameters $H_0=50\; {\rm km\; s^{-1}\; Mpc^{-1}}$ 
and $q_0=0.5$ for all cosmological calculations.

\section{Results} 

Table \ref{fit_res} gives the results of the two-dimensional model
fitting. The quality of the fits is 
generally acceptable for both the de Vaucouleurs and disk models.
The differences between the results of the two observations 
give a measure for the systematical errors, the main source of which 
are the uncertainties of the PSF determination.

The main objective of the H band imaging observations was to determine any
decentering of the nuclear point source with respect to the galaxy.
The best fit offset values as given in columns $\Delta\alpha$ and
$\Delta\delta$ of Table \ref{fit_res} show that the position
of the nucleus is in good agreement with the center of the galaxy.
In the 31 Jan observation the positional deviation is $\sim 0.05$ arcsec,
which is somewhat larger than the statistical (1$\sigma$) error of
0.015 arcsec. This deviation must be attributed to the systematic
effects of the inaccurate PSF model, as the fit to the second
observation results in very good alignment of nucleus and 
galaxy with an offset of only 0.007 arcsec.
We therefore conclude, that the decentering of the BL Lac nucleus
is less than 0.05 arcsec. 

We do not find any evidence for multiple images 
or any other deviation of the central source from a point source image.
It is therefore unlikely that the image of PKS~1413+135 is affected
by gravitational macrolensing. 

\begin{figure*}
\par\centerline{\psfig{figure=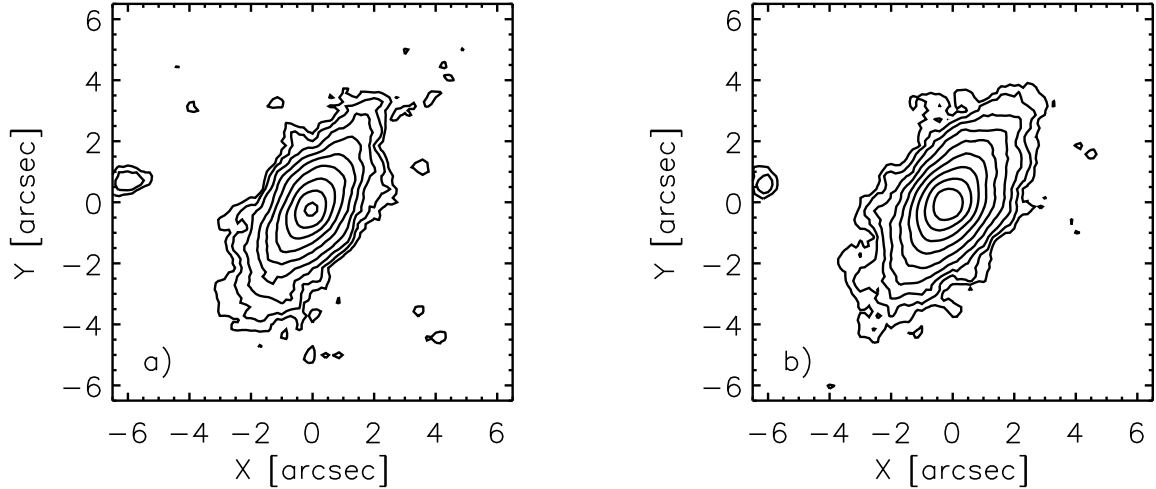,width=17truecm}}
\caption{\label{contours}H band surface brightness contours of the
galaxy associated with PKS~1413+135 after subtracting the nuclear
point source. Contours start at 21.5 mag arcsec$^-2$ and are incremented
in 0.5 mag steps. a) Observation from 31-Jan-1998, b) observation from
1-Feb-1998. North is up and east to the left.}
\end{figure*}

The morphology and magnitude of the galaxy are rather poorly
constrained as in H band the BL Lac nucleus dominates the central 
regions of the image and the parameters of the galaxy  critically 
depend on the subtraction of the point source. 
The best fit scale lengths (1.6-2.0 arcsec) and ellipticities (0.61-0.74)
agree with the values  measured at optical wavelengths 
(\ncite{imh91}, \ncite{imh94}).
The surface brightness
contours of the galaxy after removal of the best fit point source
are shown in Fig. \ref{contours}.
The outer isophotes show irregular features 
looking not too different from  box- or peanut-shaped isophotes
(see e.g. \ncite{Shaw} for examples).
Peanut shaped bulges in edge on galaxies are commonly regarded as evidence
of a buckled central bar. The bar may buckle
spontaneously
(e.g. \ncite{Raha}, \ncite{KM})
or the buckling may have been
caused by accretion of a low mass comanpanion \cite{Mihos}.

The presence of bars in spiral galaxies is thought to enhance fuelling
into any central black hole \cite{Shlosman} 
which is consistent with  the 
supposition that the BL Lac core in PKS~1413+135 actually is located
in the apparent parent galaxy.

\begin{table}  
\caption{\label{fit_res} Best fit models}
\begin{tabular}{lccccrrc}
Date & $m_{\rm gal}$ & $R_{\rm s}['']$ & $m_{\rm nuc}$ & 
$\Delta\alpha['']$ & $\Delta\delta['']$ & $\chi^2_{\rm red}$\\[10pt] 
\multicolumn{7}{c}{de Vaucouleurs model}\\[10pt]
31/01/98 & 15.75 & 2.00  & 15.69 & 0.016 & 0.048 & 1.10\\ 
01/02/98 & 15.37 & 1.68 & 15.87 &-0.005 &-0.005 & 0.73\\[10pt]   
\multicolumn{7}{c}{disk model}\\[10pt]
31/01/98 & 16.16 & 1.67 & 15.58 & 0.005 & 0.048 & 0.98\\  
01/02/98 & 15.79 & 1.58 & 15.65 &-0.004 &-0.006 & 1.20\\  
\end{tabular}
\end{table}

\section{Discussion}
\label{Disc}

\begin{figure}
\par\centerline{\psfig{figure=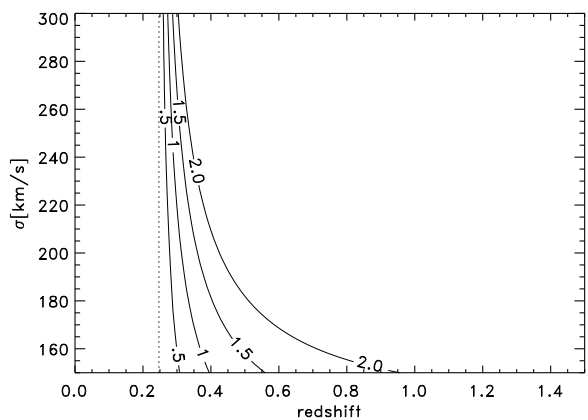,width=8.5truecm}}
\caption{\label{z_sigma} Limitations on lens properties and source
redshift for a lensing system without multiple images.  
The graph shows contours of minimum core radius
of a lensing galaxy at z=0.247 as a function of $\sigma_{\rm vel}$
and source redshift. For a given core radius and velocity
dispersion the graph can be used to find the upper limit of
the source redshift. 
}
\end{figure}

Our H band images clearly show that the position of PKS~1413+135 
is centered within $<0.05$ arcsec of its  putative host galaxy.
This corresponds to $<240$ pc in the frame of the galaxy at $z=0.247$. 
The new data show that a chance alignment of the galaxy and the 
BL Lac in the background is extremely unlikely.
Our new images reduce the upper limit on any separation of the BL
Lac core from the centre of the galaxy by at least a factor of 2
compared to the earlier observations of \scite{Stocke92} and hence
reduce the probability that the BL Lac is a background object by
at least a factor of 4. 
\scite{Stocke92} argue that microlensing by the stars in the 
foreground galaxy might increase this probability as the background
object might belong to an intrinsically  much fainter and hence more 
numerous population. However, gravitational macrolensing cannot be
switched off and its expected effects inflict stringent constraints on
the configuration of source and lens. 
No evidence for lensing effects (e.g. multiple images) is found in our 
H band images nor in VLBA data obtained by \scite{Perlman}. 

\scite{Narayan} developed a method to derive a 
lower limit for the core radius of a lensing galaxy, if there are
no multiple images of the lensed object observed. 
The limiting core radius of the lensing galaxy, modeled as an
isothermal sphere, depends on
the redshifts of lens and source as well as on 
the velocity dispersion $\sigma$ in the  lensing galaxy. 
\scite{Narayan} show that a lower limit
on the core radius can still be given in the presence of 
perturbations due to an asymmetric mass distribution in the lensing
galaxy  or due to contributions to the lensing from a
cluster potential or other nearby galaxies.   

In the case of PKS 1413+135 only one redshift, that of the spiral
galaxy, is known. Therefore we have used the method of
Narayan\&Schneider to constrain the redshift 
of any lensed object as a function of velocity dispersion 
and core radius of the galaxy.
The results are shown as contours of the minimum core radius in
the redshift-$\sigma$ plane (Fig \ref{z_sigma}). For any assumed core 
radius only the area to the left of the corresponding contour is allowed,   
if multiple imaging is to be avoided.
Even for very conservative assumptions on the bulge of the galaxy
($r_{\rm core}$=2 kpc, $\sigma_{\rm vel}$=180 km s$^{-1}$) the redshift
of the source cannot be larger than z=0.5.  
Among the possible candidates for radio sources being lensed  by the 
spiral galaxy, radio galaxies are the most abundant objects.
Assuming a 10-fold flux amplification due to microlensing, the 
limiting flux density of the candidate source is 70 mJy at 2.7 GHz.
According to the local luminosity functions of FR I and FR II radio 
galaxies (\ncite{Urry}, \ncite{Padovani}) the surface density
of radio galaxies with redshifts $0.25<z<0.5$ and exceeding the above
flux limit is 0.02 ${\rm deg}^{-1}$. The resulting probability of a chance
alignment within 0.05 arcsec  is $\sim 10^{-11}$ per galaxy. 
The total number of galaxies brighter than the putative lens 
($m_{\rm B}=20.6$) is $\sim 3\cdot10^7$ (e.g. Heydon-Dumbleton et
al. 1989), hence the probability 
of finding one microlensing system like  PKS 1413+135 is only 
$3\cdot10^{-4}$. We stress that removing the redshift limit 
imposed on the population of radio sources  
does not increase the probability of a chance alignment significantly.
We therefore conclude that the disk galaxy must be the host of
PKS~1413+135. 

However, the existence of a BL Lac object within a spiral galaxy raises
several problems:  

\begin{enumerate}

\item Absorption of isotropic radiation from the AGN nucleus is 
likely to give rise
to thermal emission due to re-radiation from dust at far-infrared 
wavelengths. In a gas rich spiral galaxy
broad and narrow emission line  regions would be expected.   
However, the far-infrared emission of PKS~1413+135 \cite{Impey}
is probably mostly non-thermal, as significant variability has been
detected. Nevertheless, a weak thermal continuum might be present. 
No emission lines have been detected in near-infrared spectra 
of PKS~1413+135, the upper limit for the luminosity of the Paschen 
$\alpha$ line is $\sim 10^7 L_{\sun}$ \cite{Perlman}.    
The emission line luminosity is clearly lower than in Seyfert galaxies
or FR II galaxies  of compareable X-ray luminosity, but
is more compatible with the line luminosities of FR I radio galaxies.
\scite{Baum} argue that the lower emission line luminosities 
in FR I sources is not due to a lack of surrounding gas to be ionised,
but is related to the absence of a strong isotropically emitted UV and
X-ray continuum from the central engine. If the nucleus of
PKS~1413+135 is similar to a FR I nucleus and the observed emission  
is sufficiently beamed, i.e. the total luminosity
is low, the lack of  reprocessed radiation can be understood.

\item If the jet of PKS~1413+135 is pointed towards us, as expected 
from the BL Lacertae properties of the source, it is oriented 
close to the disk plane of the spiral galaxy.
The drastic misalignment of the jet axis with respect  to the 
spin axis of the stellar disk is at first surprising. On the other
hand statistical investigations of the radio axes in  Seyfert galaxies
revealed that the orientations of accretion disks and galactic disks  
generally are not correlated (e.g. \ncite{Ulvestad}, \ncite{Clarke}.   

\item 
If the jet is pointed towards us and originates in a nearly edge-on
galaxy, an obvious question is how the jet of an active nucleus could make its
way through the disc of a large, dusty spiral galaxy. 
\scite{Perlman} report that the jets of PKS 1413+135 are very
compact with an overall projected length of less than 1 kpc. Their detection of
jet and counterjet suggests that although the core component is beamed,
the viewing angle of the jet is too large for significant Doppler
boosting. The jet is obviously bent away from the line of
sight and intrinsically very compact. Hence it might well be confined 
to the central, bulge dominated, regions of the galaxy.
\scite{Perlman} point out the similarity of the radio  morphology with 
CSOs (compact steep objects, \ncite{Wilkinson}). The compactness
of the jets might be caused by the comparatively high density of 
the interstellar gas in the spiral galaxy or, as \scite{Perlman} suggest,
may be an indication that PKS~1415+135 is a very young radio galaxy.

\item A fundamental difference between radio loud and radio quiet 
AGN is the fact that with very few exceptions radio quiet AGN are
found in spiral host galaxies while radio galaxies are
associated with elliptical galaxies. It is generally believed that mergers 
of galaxies play an important role in the creation of radio loud AGN.
\scite{Wilson} propose that the merger of two spiral galaxies  
(or one spiral galaxy and one elliptical galaxy) each containing a 
supermassive black hole lead to the formation of radio loud AGN
powered by a rotating black hole inside an elliptical galaxy.  
\scite{Colina} find evidence for recent mergers 
or interactions in a large fraction of FR I host galaxies and propose
that the FR I galaxies are the product of two merging ellipticals.
The spiral host galaxy of PKS~1413+135 is clearly an exception that
does not readily fit into the standard models of radio loud AGN. 
However, the FR I radio source (0313-192) discovered in a disk
dominated host galaxy \cite{Ledlow} might be an unbeamed 
counterpart of PKS~1413+135.  
terms of morphology and luminosity this galaxy is similar to the 
PKS~1413+135 host and a prominent dust lane has been found in both objects.
The radio galaxy 0313-192 might be the same intrinsic type    
as PKS~1413+135 and the physical problems  related with the
development of a powerful jet in a gas rich disk galaxy obviously
are the same for both objects. 
Our detection of a peanut-shaped bulge in the host galaxy of PKS~1413+135
indicates that the  galaxy contains a central bar.
Some models of X-shaped and peanut-shaped galaxies involve 
the accretion of small satellites leading to the buckling the bar
\cite{Mihos}. 
The host galaxy of PKS~1413+135 is not likely to be a product
of a major merger event. However, the presence of a central bar and 
the possible accretion of satellite galaxies are  among the
prerequisites for an AGN in the centre of the galaxy.

\end{enumerate}

We conclude that despite the problems raised by the supposition 
of a BL Lac nucleus in a spiral host, the spiral galaxy is 
almost certainly the host galaxy of PKS~1413+135. The 
hypothesis that the radio loud AGN is background to the galaxy 
can be ruled out for statistical reasons, as we show that 
lensing or microlensing by the galaxy does not significantly
increase the probability of the  close alignment of  galaxy
and AGN.


\section*{Acknowledgments}

Thanks to Mike Merrifield for discussions about peanut-shaped galaxies.
The United Kingdom Infrared Telescope (UKIRT) is operated by the 
Joint Astronomy Centre on behalf of the UK Particle Physics and
Astronomy
Research Council. {\tt IRAF} is distributed by the National Optical
Astronomy Observatories, which are operated by the Association of
Universities for Research in Astronomy, Inc., under cooperative
agreement
with the National Science Foundation.

\end{document}